\documentclass[twocolumn,aps,showpacs,floatfix,prc]{revtex4}
\usepackage[dvips]{epsfig}
\begin{document}
%\draft
\title{Relativistic Feynman-Metropolis-Teller Equation of State\\ for White Dwarfs in presence of Magnetic Field}

\author{ Sujan Kumar Roy$^{1*\S}$, Somnath Mukhopadhyay$^{2*\S\dagger}$, Joydev Lahiri$^{3*}$ and D.N. Basu$^{4*\S}$ }

\affiliation{$^*$ Variable  Energy  Cyclotron  Centre, 1/AF Bidhan Nagar, Kolkata 700 064, India }
\affiliation{$^{\S}$ Homi Bhabha National Institute, Training School Complex, Anushakti Nagar, Mumbai 400085}
\address{ $^\dagger$ BITS Pilani, Hyderabad Campus, Jawahar Nagar, Kapra Mandal, Medchal District 500078
Telangana, India}

\email[E-mail 1: ]{sujan.kr@vecc.gov.in}
\email[E-mail 2: ]{somnathm@hyderabad.bits-pilani.ac.in}
\email[E-mail 2: ]{joy@vecc.gov.in}
\email[E-mail 3: ]{dnb@vecc.gov.in}
\date{\today }

\begin{abstract}

    The relativistic Feynman-Metropolis-Teller treatment of compressed atoms is extended to treat magnetized matter. Each atomic configuration is confined by a Wigner-Seitz cell and is characterized by a positive electron Fermi energy which varies insignificantly with the magnetic field. In the relativistic treatment the limiting configuration is reached when the Wigner-Seitz cell radius equals the radius of the nucleus with a maximum value of the electron Fermi energy which can not be attained in presence of magnetic field due to the effect of Landau quantization of electrons within the Wigner-Seitz cell. This treatment is implemented to develop the Equation of State for magnetized White Dwarf stars in presence of Coulomb screening. The mass-radius relations for magnetized White Dwarfs are obtained by solving the general relativistic hydrostatic equilibrium equations using Schwarzschild metric description suitable for non rotating and slowly rotating stars. The explicit effects of the magnetic energy density and pressure contributed by a density-dependent magnetic field are included to find the stable configurations of magnetized Super-Chandrasekhar White Dwarfs.     
%\vskip 0.2cm


\end{abstract}


\maketitle

\noindent
\section{Introduction}
\label{section1}

    Compact stars are home to the strongest magnetic fields in the universe. Anomalous X-ray Pulsars (AXPs) and Soft Gamma Repeaters (SGRs), which are certain classes of neutron stars called magnetars, are observed to have surface magnetic fields   $\sim 10^{15}$ gauss. Moreover, a strong magnetic field of similar magnitude has been observed at the jet base of a supermassive black hole PKS 1830-211 \cite{Vi15}. Several magnetized white dwarfs have also been observed with surface field strengths ranging from $\sim 10^{6}-10^{9}$ gauss \cite{Ke13,Ke15,Fe15}. Such strong magnetic fields can drastically modify the Equation of State (EoS) of a compact star, its structure and stability.
    
    A stellar magnetic field is a magnetic field generated by the motion of conductive plasma inside a star. This motion is created through convection, which is a form of energy transport involving the physical movement of material. A localized magnetic field exerts a force on the plasma, effectively increasing the pressure without a comparable gain in density. As a result, the magnetized region rises relative to the remainder of the plasma, until it reaches the star's photosphere. This creates star spots on the surface, and the related phenomenon of coronal loops. Stellar magnetic fields, according to solar dynamo theory, are caused within the convective zone of the star. The convective circulation of the conducting plasma functions like a dynamo. This activity destroys the star's primordial magnetic field then generates a dipolar magnetic field. As the star undergoes differential rotation rotating at different rates for various latitudes the magnetism is wound into a toroidal field of "flux ropes" that become wrapped around the star. The fields can become highly concentrated, producing activity when they emerge on the surface.

    The magnetic fields of all celestial bodies are often aligned with the direction of rotation, with notable exceptions such as certain pulsars. Another feature of this dynamo model is that the currents are AC rather than DC. Their direction, and thus the direction of the magnetic field they generate, alternates more or less periodically, changing amplitude and reversing direction, although still more or less aligned with the axis of rotation. Such magnetized stars when collapse the magnetic fields of the progenitors are trapped within new compact stars. The origin of high magnetic field of the collapsed stars is due to the flux conservation in the new configuration. 
    
    Recent studies \cite{Das12,Das13,Ni14,Mu17} have shown the existence of magnetized white dwarfs with masses significantly greater than the traditional Chandrasekhar's limit \cite{Ch35},also known as Super-Chandrasekhar white dwarfs to account for the exceptionally high luminosities of the Type Ia supernovae, e.g. SN2006gz, SN2007if, SN2009dc, SN2003fg \cite{Ho06,Sc10,Hi07,Ya09,Si11}. In these studies, the electron gas in the EoS is taken to be free, relativistic and Landau quantized in a strong magnetic field. Hence the masses and radii are independent of the composition of the white dwarf. Coulomb corrections to the EoS within the relativistic Feynman-Metropolis-Teller (FMT) formalism of compressed atoms at zero \cite{Ro11} and finite temperature \cite{Ca14} in the absence of magnetic field have been explored previously. In the present work, we extend the relativistic FMT approach to include the effects of Landau quantization of the electrons in a magnetic field and thus obtain the masses and radii of helium, carbon, oxygen and iron Super-Chandrasekhar white dwarfs .

\noindent
\section{Landau quantization for electrons in magnetic field }
\label{section2}

    We consider a completely degenerate relativistic electron gas at zero temperature in a strong magnetic field. At this stage, any form of interactions with or among electrons is not considered. Electrons, being charged particles, occupy Landau quantized states in a magnetic field. This effect causes changes in the EoS of a non-interacting degenerate electron gas. It should be emphasized that protons also, being charged particles, are Landau quantized. As the proton is $\sim 1837$ times heavier than electron, its cyclotron energy is $\sim 1837$ times smaller than that of the electron for the same magnetic field, and hence can be neglected.

    In order to calculate the thermodynamic quantities like the energy density and pressure of an electron gas in a magnetic field, we need to know the density of states and the dispersion relation. The quantum mechanics of a charged particle in a magnetic field is presented in many texts (e.g. Sokolov and Ternov (1968) \cite{ST68}, Landau and Lifshitz (1977) \cite{LL77}, Canuto and Ventura (1977) \cite{CV77}  M\'esz\'aros (1992) \cite{Me92}). Here we summarize the basics needed for our later discussion. Let us first consider the motion of a charged particle (charge $q$ and mass $m_e$) in a uniform magnetic field $B$ assumed to be along the z-axis. In classical physics, the particle gyrates in a circular orbit with radius and angular frequency (cyclotron frequency) given by
    
\begin{equation}
 r_c =\frac{m_ecv_{\perp}}{qB};    ~~~~    \omega_c=\frac{qB}{m_ec}
\label{seqn3}
\end{equation}    
where $v_{\perp}$ is the velocity perpendicular to the magnetic field. The hamiltonian of the system is given by  

\begin{equation}
H =\frac{1}{2m_e} \Big(\vec p - \frac{q\vec A}{c}\Big)^2
\label{seqn4}
\end{equation} 
where $\vec B = \nabla \times \vec A$ with $\vec A$ being the electromagnetic vector potential. To have magnetic field in z-direction with magnitude $B$ one can have

\begin{equation}
 \vec A = \bordermatrix{ &\cr 
                        &0 \cr
                        &Bx \cr
                        &0\cr}                      
\label{seqn5}
\end{equation}
and therefore

\begin{equation}
H =\frac{1}{2m_e} [p_x^2 + \Big(p_y - \frac{qBx}{c}\Big)^2 + p_z^2]
\label{seqn6}
\end{equation}
The operator $\hat p_y$ commutes with this hamiltonian since the operator $y$ is absent. Thus operator $\hat p_y$ can be replaced by its eigenvalue $\hbar k_y$. Using cyclotron frequency $\omega_c=\frac{qB}{m_ec}$ one obtains 

\begin{equation}
H =\frac{p_x^2}{2m_e} + \frac{1}{2}m_e\omega_c^2\Big(x - \frac{\hbar k_y}{m_e\omega_c}\Big)^2 + \frac{p_z^2}{2m_e},
\label{seqn7}
\end{equation}
the first two terms of which is exactly the quantum harmonic oscillator with the minimum of the potential shifted in co-ordinate space by $x_0=\frac{\hbar k_y}{m_e\omega_c}$. Noting that translating harmonic oscillator potential does not affect the energies, energy eigenvalues can be given by

\begin{equation}
E_{n ,p_z} =(n+\frac{1}{2})\hbar \omega_c + \frac{p_z^2}{2m_e}, ~~~~n=0, 1, 2 . . . . 
\label{seqn8}
\end{equation}
The energy does not depend on the quantum number $k_y$, so there will be degeneracies. Each set of wave functions with same value of $n$ is called a Landau Level. Each Landau level is degenerate due to the second quantum number $k_y$. If periodic boundary condition is assumed $k_y$ can take values $k_y=\frac{2\pi N}{l_y}$ where $N$ is another integer and $l_x,l_y,l_z$ being the dimensions of the system. The allowed values of $N$ are further restricted by the condition that the centre of the force of the oscillator $x_0$ must physically lie within the system, $0\le x_0 \le l_x$ which implies $0\le N \le \frac{l_x l_y m_e\omega_c }{2\pi\hbar}=\frac{qB l_x l_y}{hc}$. Hence for electrons with spin $s$, rest mass $m_e$ and charge $q=-e$, the maximum number of particles per Landau level per unit area is $\frac{eB(2s+1)}{hc}$. On solving Schr\"odinger's equation for electrons with spin in an external magnetic field in z-direction which is uniform and static, Eq.(\ref{seqn8}) modifies to 

\begin{equation}
E_{\nu ,p_z}=\nu\hbar \omega_c + \frac{p_z^2}{2m_e}, ~~~~ \nu=n+\frac{1}{2}+s_z.
\label{seqn9}
\end{equation}
Clearly for the lowest Landau level ($\nu=0$) the spin degeneracy $g_\nu=1$ (since only $n=0$, $s_z=-\frac{1}{2}$ is allowed) and for all other higher Landau levels ($\nu\neq0$), $g_\nu=2$ (for $s_z=\pm\frac{1}{2}$).

    For extremely strong magnetic fields such that $\hbar \omega_c \geq m_ec^2$ the motion perpendicular to the magnetic field still remains quantized but becomes relativistic. The solution of the Dirac equation in a constant magnetic field \cite{Lai91} is given by the energy eigenvalues
    
\begin{equation}
E_{\nu ,p_z}=\left[p_z^2c^2+m_e^2c^4\left(1+2\nu B_D\right)\right]^\frac{1}{2}
\label{seqn10}
\end{equation}    
where the dimensionless magnetic field defined as $B_D=B/B_c$ is introduced with $B_c$ given by $\hbar\omega_c=\hbar\frac{eB_c}{m_ec}=m_ec^2 \Rightarrow B_c=\frac{m_e^2c^3}{e\hbar}=4.414\times 10^{13}$ gauss. Obviously, the density of states in presence of magnetic field gets modified to 

\begin{equation}
\sum\limits_{\nu }\frac{2eB}{hc}g_{\nu}\int \frac{dp_z}{h}
\label{seqn11}
\end{equation}
where the sum is on all Landau levels $\nu$. At zero temperature the number density of electrons is given by

\begin{equation}
n_e=\sum\limits_{\nu =0}^{\nu_m} \frac{2eB}{h^2c} g_{\nu} \int_0^{p_F(\nu )} dp_z = \sum\limits_{\nu =0}^{\nu_m}\frac{2eB}{h^2c}g_{\nu}p_F(\nu)
\label{seqn12}
\end{equation}
where $p_F(\nu )$ is the Fermi momentum in the $\nu $th Landau level and $\nu_m$ is the upper limit of the Landau level summation. The Fermi energy $E_F$ of the $\nu $th Landau level is given by

\begin{equation}
E_F^2=p_F^2(\nu)c^2+m_e^2c^4\left(1+2\nu B_D\right)
\label{seqn13}
\end{equation}
and $\nu_m$ can be found from the condition $[p_F(\nu)]^2 \geq 0$ or

\begin{equation}
\nu \leq \frac{\epsilon_F^2-1}{2B_D} ~~\Rightarrow ~~ \nu_m =\frac{\epsilon_{F max}^2-1}{2B_D},
\label{seqn14}
\end{equation}     
where $\epsilon_F=\frac{E_F}{m_ec^2}$ is the dimensionless Fermi energy and $\epsilon_{F max}=\frac{E_{F max}}{m_ec^2}$ the dimensionless maximum Fermi energy of a system for a given $B_D$ and $\nu_m$. Obviously, very small $B_D$ corresponds to large number of Landau levels tending to the non-magnetic case. $\nu_m$ is taken to be the nearest lowest integer. In terms of the dimensionless Fermi momentum $x_F(\nu)=\frac{p_F(\nu )}{m_ec}$, Eq.(\ref{seqn12}) and Eq.(\ref{seqn13}) may be written as

\begin{equation}
n_e=\frac{2B_D}{(2\pi )^2\lambda_e^3}\sum\limits_{\nu =0}^{\nu_m} g_{\nu}x_F(\nu )
\label{seqn15}
\end{equation}
where $\lambda_e=\frac{\hbar}{m_e c}$ is the Compton wavelength for electron and

\begin{equation}
\epsilon_F=\left[x^2_F(\nu )+1+2\nu B_D\right]^\frac{1}{2}
\label{seqn16}
\end{equation}
or

\begin{equation}
x_F(\nu )=\left[\epsilon_F^2-(1+2\nu B_D)\right]^\frac{1}{2}.
\label{seqn17}
\end{equation}   
The electron energy density is given by

\begin{eqnarray}
\varepsilon_e&&=\frac{2B_D}{(2\pi )^2\lambda_e^3}\sum\limits_{\nu =0}^{\nu_m} g_{\nu }\int_0^{x_F(\nu )}E_{\nu ,p_z}d\left(\frac{p_z}{m_ec}\right) \\
&&=\frac{B_D}{2\pi^2}\frac{m_ec^2}{\lambda_e^3}\sum\limits_{\nu =0}^{\nu_m} g_{\nu }(1+2\nu B_D)\psi \left(\frac{x_F(\nu )}{(1+2\nu B_D)^{1/2}}\right),  \nonumber
\label{seqn18}
\end{eqnarray}
where

\begin{equation}
\psi (z)=\int_0^z(1+y^2)^{1/2}dy=\frac{1}{2}[z\sqrt{1+z^2}+\ln(z+\sqrt{1+z^2})]
\label{seqn19}
\end{equation}
The pressure of the electron gas is given by

\begin{eqnarray}
P_e&&=n_e^2\frac{d}{dn_e}\left(\frac{\varepsilon_e}{n_e}\right)= n_eE_F -\varepsilon_e \\
&&=\frac{B_D}{2\pi^2}\frac{m_ec^2}{\lambda_e^3}\sum\limits_{\nu =0}^{\nu_m} g_{\nu }(1+2\nu B_D)\eta \left(\frac{x_F(\nu )}{(1+2\nu B_D)^{1/2}}\right), \nonumber
\label{seqn20}
\end{eqnarray}
where

\begin{equation}
\eta (z)=z\sqrt{1+z^2}-\psi (z) =\frac{1}{2}[z\sqrt{1+z^2}-\ln(z+\sqrt{1+z^2})].
\label{seqn21}
\end{equation}

\noindent
\section{The Relativistic Feynman-Metropolis-Teller treatment in presence of magnetic field} 
\label{section3}
    
    For the extension of FMT formalism in presence of magnetic field, a compressed atom as a Wigner-Seitz cell consisting of a finite size nucleus with mass number $A$ and atomic number $Z$ at the centre of the cell and completely degenerate relativistic electron gas embedded in a strong magnetic field is considered. As described in the preceding section, electrons being charged particles occupy Landau quantized states in a magnetic field $B$ and the maximum number of particles per Landau level per unit area is $\frac{eB(2s+1)}{hc}$. The Fermi energy $E_F$ of the $\nu $th Landau level in the external magnetic field $B$ in z-direction and under the influence of Coulomb potential $V(r)$ can now be given by 
    
\vspace{0.cm}
\begin{equation}
\vspace{0.cm}
E_F=\left[p_F^2(\nu)c^2+m_e^2c^4\left(1+2\nu B_D\right)\right]^\frac{1}{2}-m_ec^2-eV(r)
\label{seqn22}
\vspace{0.0cm}
\end{equation}
\noindent
where the Fermi Energy is constant over a compressed Wigner-Seitz cell. A constant number density distribution of protons $n_p(r)$ is assumed which is confined within nuclear radius $R_c=r_0A^\frac{1}{3}=\Delta\lambda_\pi Z^\frac{1}{3}$ with $\lambda_\pi=\frac{\hbar}{m_\pi c}$ being the Compton wavelength for pion of rest mass $m_\pi$ and hence can be given by $n_p(r) = \frac{Z}{\frac{4\pi}{3}R_c^3}~\theta(R_c-r)$. Defining $\widehat{V}(r)=eV(r)+E_F$, from Eq.(\ref{seqn22}) $p_F(\nu)=\sqrt{\widehat{V}^2+2\widehat{V}m_ec^2-2\nu B_Dm_e^2c^4}$ and from the condition $p_F^2\geq0$, the upper limit of Landau level  $\nu_m$ can be obtained as $\nu_m=\frac{\widehat{V}^2+2\widehat{V}m_ec^2}{2B_Dm_e^2c^4}$.

    Using Landau quantization, the number density of electrons under the influence of Coulomb screening is, therefore, given by

\vspace{0.cm}
\begin{eqnarray}
\vspace{0.cm}
n_e(r)=&&\frac{2B_D}{(2\pi )^2\lambda_e^3}\sum\limits_{\nu =0}^{\nu_m} g_{\nu}x_F(\nu) \nonumber \\
{\rm where}~p_F(\nu)c=&&\left[\left(\widehat{V}^2+2m_ec^2\widehat{V}\right) \left(1-\frac{\nu}{\nu_m}\right)\right]^\frac{1}{2} 
\label{seqn23}
\vspace{0.cm}
\end{eqnarray}
\noindent
with $x_F(\nu)=\frac{p_F(\nu )c}{m_ec^2}$ as before. The overall Coulomb potential $V(r)$ can be obtained by solving the Poisson equation

\vspace{0.cm}
\begin{eqnarray}
\vspace{0.cm}
\nabla^2V(r)=-4\pi e[n_p(r)-n_e(r)] \nonumber\\
\Rightarrow \nabla^2\widehat{V}(r)=-4\pi e^2[n_p(r)-n_e(r)] 
\label{seqn24}
\vspace{0.cm}
\end{eqnarray}
\noindent

    Introducing the dimensionless quantities $x=\frac{r}{\lambda_\pi}$ and $\chi(x)= r\frac{\widehat{V}(r)}{\hbar c}=x\frac{\widehat{V}(x)}{m_\pi c^2}$, the Eq.(\ref{seqn24}) can be rewritten as 
    
\begin{eqnarray}
\frac{1}{x}\frac{d^2\chi(x)}{dx^2}= -\frac{3\alpha \theta(x_c-x)}{\Delta^3}\nonumber\\
+\frac{2 e^2B_D}{\pi \hbar c}\left(\frac{m_\pi}{m_e}\right)\left(\frac{\lambda_\pi}{\lambda_e}\right)^3 \times \nonumber\\ 
\sum\limits_{\nu =0}^{\nu_m} g_{\nu}\left[\left(1-\frac{\nu}{\nu_m}\right)\left\{\left(\frac{\chi}{x}\right)^2+2\left(\frac{m_e}{m_\pi}\right)\left(\frac{\chi}{x}\right)\right\}\right]^\frac{1}{2}
\label{seqn25}
\vspace{0.cm}
\end{eqnarray}
\noindent
where $\alpha$ is the fine structure constant. The electrostatic potential and the number density distribution of electrons can be obtained by solving Eq.(\ref{seqn25}) with the boundary conditions $\chi(0)=0$ and $\chi\prime(x_{WS})=\chi(x_{WS})/x_{WS}$ where $x_{WS}$ is the dimensionless radius of the Wigner-Seitz cell. The first boundary condition follows from the fact that potential can not become infinite at the centre of the cell, while the second boundary condition implies the charge neutrality which must ensure charge number conservation $Z=\int_0^{r_{WS}} 4\pi n_e(r) r^2 dr$ as well as $\chi(x_{WS}) \geq 0$ that preserves the basic assumption of FMT approach that a compressed Wigner-Seitz cell is characterized by a positive Fermi Energy.

\noindent
\section{ The equation of state }
\label{section4}

    Once the relativistic FMT is solved for a compressed atom in presence of magnetic field, the EoS can be constructed. The kinetic energy density including the electronic rest mass within the Wigner-Seitz cell is given by   
    
\vspace{0.0cm}
\begin{equation}
\vspace{0.cm}
\varepsilon_k (x)=\frac{B_Dm_ec^2}{2\pi^2\lambda_e^3} \sum\limits_{\nu =0}^{\nu_m} g_\nu\left(1+2\nu B_D\right) 
\psi \left(\frac{x_F(\nu )}{(1+2\nu B_D)^{1/2}}\right)  \nonumber\\
\label{seqn26}
\vspace{0.cm}
\end{equation}
\noindent
at radius (dimensionless) $r~(x)$ from the centre of the cell. The dependence on $x$, unlike the previous case (section II), is due to inclusion of Coulomb screening in Eq.(\ref{seqn23}). The total kinetic energy $E_k$ of the cell excluding the electronic rest mass can be calculated as 
    
\vspace{0.cm}
\begin{equation}
\vspace{0.cm}
E_k=4\pi\lambda_\pi^3\int_{x_c}^{x_{WS}} x^2[\varepsilon_k (x)-n_e(x)m_ec^2]dx
\label{seqn27}
\vspace{0.0cm}
\end{equation}
\noindent
where $x_c=\frac{R_c}{\lambda_\pi}$ is the dimensionless radius of the nucleus resting at the centre of the cell. As the calculations will be performed up to the onset of inverse $\beta$ decay, the electrons are absent inside the nucleus and the integration is performed from $x_c$.   

    The total potential energy $E_c$ of the cell can be evaluated using
    
\vspace{0.cm}
\begin{eqnarray}
\vspace{0.cm}
E_c=4\pi\lambda_\pi^3\int_0^{x_{WS}} x^2[n_p(x)-n_e(x)]eV(x)dx \nonumber \\
\Rightarrow E_c=-4\pi\lambda_\pi^3\int_{x_c}^{x_{WS}} x^2n_e(x)eV(x)dx \nonumber \\
=-4\pi\lambda_\pi^3 m_\pi c^2 \int_{x_c}^{x_{WS}} xn_e(x)\chi(x)dx +E_FZ
\label{seqn28}
\vspace{0.0cm}
\end{eqnarray}
\noindent
where the lower limit of the integration is changed to $x_c$ since in the forthcoming calculations, while calculating energy density, measured atomic masses will be used implying that electronic rest mass and Coulomb energy contribution due to nucleus will be accounted and the term $n_p$ is dropped since protons are not present beyond $x_c$. It is, therefore, obvious that in Eq.(\ref{seqn27}) electronic rest mass is subtracted to avoid double counting. 

    The energy density $\varepsilon$ can now be given by

\vspace{0.0cm}
\begin{equation}
\vspace{0.cm}
\varepsilon=\frac{E_k+E_c+M(A,Z)c^2}{\frac{4\pi}{3}r_{WS}^3}+\frac{B^2}{8\pi} 
\label{seqn29}
\vspace{0.cm}
\end{equation}
\noindent    
where $M(A,Z)$ is atomic mass of the uncompressed atom and the last term accounts for the magnetic contribution to the energy density. The pressure $P$ is simply given by

\vspace{0.cm}
\begin{eqnarray}
\vspace{0.cm}
P=&&\frac{B_Dm_ec^2}{2\pi^2\lambda_e^3}\sum\limits_{\nu =0}^{\nu_m} g_\nu\left(1+2\nu B_D\right)\eta \left(\frac{x_F(\nu )}{(1+2\nu B_D)^{1/2}}\right) \nonumber\\
&&+\frac{B^2}{24\pi} 
\label{seqn30}
\vspace{0.cm}
\end{eqnarray}
\noindent
but it should be evaluated at the cell boundary and the last term accounts for the magnetic contribution to the pressure.
        
\noindent
\section{ Results and discussion }
\label{section5}

    Recently, there are some important calculations for masses and radii of magnetized white dwarfs using non-relativistic Lane-Emden equation assuming a constant magnetic field throughout which provided masses up to 2.3-2.6 $M_\odot$ \cite{Das12}, a mass significantly greater than the Chandrasekhar limit. However, because of the structure of the Lane-Emden equation, pressure arising due to constant magnetic field do not contribute while for the general relativistic Tolman-Oppenheimer-Volkoff (TOV) \cite{TOV39a,TOV39b} equation case is not the same. Moreover, the EoS needed to be fitted to a polytropic form. In order to derive a mass limit for magnetized white dwarfs (similar to the mass limit of $\sim$1.4 $M_\odot$ obtained by Chandrasekhar \cite{Ch35} for non-magnetic white dwarfs), the same authors, under certain approximations, have been able to reduce the EoS to a polytropic form with index $1+1/n=2$ for which analytic solution of Lane-Emden equation exists ($\theta(\xi)=sin\xi/\xi$ where $\rho=\rho_c\theta^n$ with $\rho$ and $\rho_c$ being density and central density, respectively) and avoiding the energy density $\varepsilon_B=\frac{B^2}{8\pi}$ and pressure $P_B=\frac{1}{3}\varepsilon_B$ arising due to magnetic field by assuming it to be constant throughout, they were able to set a mass limit of 2.58 $M_\odot$ \cite{Das13,Do14}. The masses and radii of white dwarfs have been calculated by solving the general relativistic TOV equation both for non-magnetic and magnetized white dwarfs using the exact EoS without resorting to fit it to a polytropic form \cite{Mu17}. 

    In this present work, to find out the critical mass limit for magnetized white dwarfs the general relativistic hydrostatic equilibrium equation (TOV) has been solved for the EoS obtained up to the density for onset of inverse $\beta$ decay. The critical density $\rho^{\beta,unif}_{crit}$ for the onset of inverse $\beta$ decay within the approximation of uniform electron density distribution, is given by

\begin{equation}
 \rho^{\beta,unif}_{crit} = \frac{A_r}{Z}\frac{M_u}{3\pi^2 (\hbar c)^3}\left[(\epsilon_\beta(Z))^2 + 2m_e c^2 \epsilon_\beta(Z) \right]^{3/2}
\label{seqn31}
\end{equation} 
\noindent
where $A_r$ is the measured mass in atomic mass unit, $M_u$ is the atomic mass unit expressed in MeV and $\epsilon_\beta(Z)$ is the energy to ignite the inverse $\beta$-decay. The experimental energies for inverse $\beta$-decay are listed in Table-I. The values of the critical densities $\rho^{\beta,unif}_{crit}$ have been calculated using Eq.(\ref{seqn31}) for the decays given in Table-I. The $A_r$ values used for $^4$He, $^{12}$C, $^{16}$O and $^{56}$Fe are 4.003, 12.01, 16.00 and 55.84, respectively. The EoS calculations for magnetic field $B_D$ have been performed up to maximum (critical) densities restricted by the condition that $E_F$ given by Eq.(\ref{seqn22}) reaches asymptotically the inverse $\beta$ decay threshold energy $\epsilon_\beta(Z)$. The critical densities $\rho^{\beta,relFMT}_{crit}$ for $B_D = 5B_c$ have also been provided in Table-I. The inverse $\beta$ decay threshold energies have been assumed to be independent of magnetic field. Under this assumption, the values of $\rho^{\beta,relFMT}_{crit}$ do not differ much from zero magnetic field since the change in Fermi energy $E_F$ has a weak dependence on magnetic field.       

    If a charged particle having charge $q$ is moving with velocity $v$ perpendicular to an external magnetic field $B$ the equilibrium is established through $\frac{mv^2}{r}=qvB$ i.e. the magnetic field $B=\frac{mv}{qr}$ implying higher the magnetic field lower the radius of the trajectory of the particle. This simple logic applies to the atomic configuration subjected to magnetic field as well. Hence it is expected in the light of the above logic that the radius of a particular cell will decrease with the magnetic field. As described earlier in section \ref{section3} in presence of magnetic field the relativistic FMT treatment has been applied to study the changes in the Wigner-Seitz cell and it is found that the density increases with magnetic field, though the variation of density with magnetic field is not as fast as $\rho \propto B^3$ which can be accounted for the coulomb screening considered under FMT treatment as well as the 3$-D$ configuration of Wigner-Seitz cell. On the other hand in presence of magnetic field the landau quantization in the perpendicular plane becomes effective as described in the section \ref{section2}. Hence we expect some discreteness in the electron density distribution due to the occupation of electrons in different landau levels and this is confirmed from any of the Figs.-1-8. The gap between two successive landau level is $\hbar \frac{eB}{m_ec}$, so it is expected that under very low magnetic field the landau levels are very close to each other almost resembling a continuum alike no-magnetic field situation. This can be confirmed from the Figs.7,8 where it is clear that the difference from the no-magnetic field distribution is least for the lowest magnetic field. Alongside an atom with same density characterized by same Fermi energy if treated under high magnetic field the gap between successive landau levels increases implying the lowering of $\nu_{max}$ obvious from Figs.-9,10 as well as higher the magnetic field abrupt the $\nu_{max}$ with respect to $x$ for the same reason. In FMT treatment each compressed atomic configuration is characterized by a constant Fermi energy. But as an electron closer to the nucleus having tight bound to nucleus will carry more and more kinetic energy implying for a particular magnetic field the number of landau quantum number should be more near the nucleus to accommodate all the electrons carrying higher kinetic energy. The dilute and distant electrons are more vulnerable to the effects of magnetic field. Hence higher the quantization effect and lower the $\nu_{max}$ towards the surface of the cell. These are depicted in each of the curves in Figs.-9,10. Considering all these facts it can be inferred that most compressed atoms are least affected by the magnetic field can be seen from the comparisons of the Figs.-1,2, Figs.-3,4, Figs.-5,6, Figs.-7,8 that the differences in the distributions are reduced in the higher density domain. As an atom treated under different magnetic fields, higher the magnetic field more will be the compression subjected to the conservation of electron number and thus adopting the compression $\nu_{max}$ values are expected to be more or less same as can be seen from the comparison between 1$B_c$ curve of Fig.-9 and 10$B_c$ curve of Fig.-10. 

\begin{figure}[ht!]
\vspace{0.0cm}
\eject\centerline{\epsfig{file=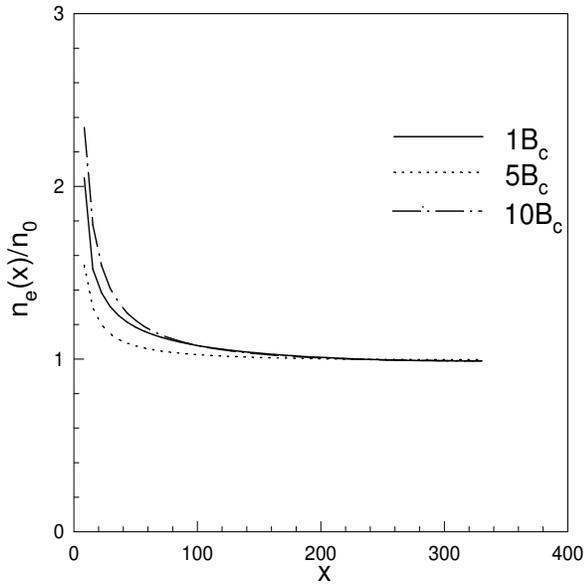,height=7.7cm,width=7.7cm}}
\caption
{Plots for $n_e({\rm x})/n_0$ as functions of dimensionless radial co-ordinate ${\rm x}$ at density $1.53\times 10^7$ gm cm$^{-3}$ for $^4$He.}
\label{fig1}
\vspace{4.0cm}
\end{figure}
\noindent 

\begin{figure}[hb!]
\vspace{0.0cm}
\eject\centerline{\epsfig{file=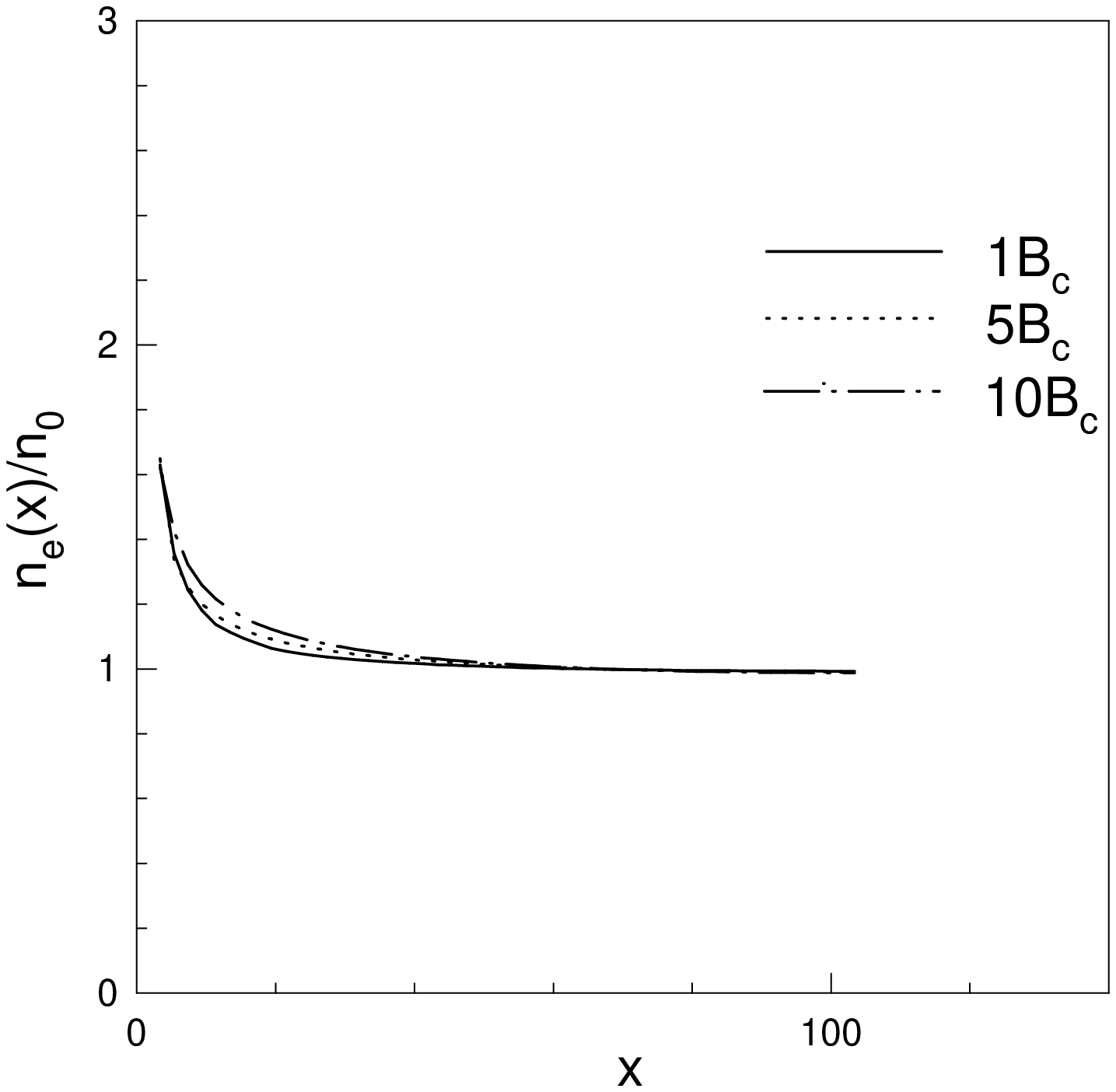,height=7.7cm,width=7.7cm}}
\caption
{Plots for $n_e({\rm x})/n_0$ as functions of dimensionless radial co-ordinate ${\rm x}$ at density $4.96\times 10^8$ gm cm$^{-3}$ for $^4$He.}
\label{fig2}
\vspace{0.0cm}
\end{figure}
\noindent 

\begin{figure}[ht!]
\vspace{0.0cm}
\eject\centerline{\epsfig{file=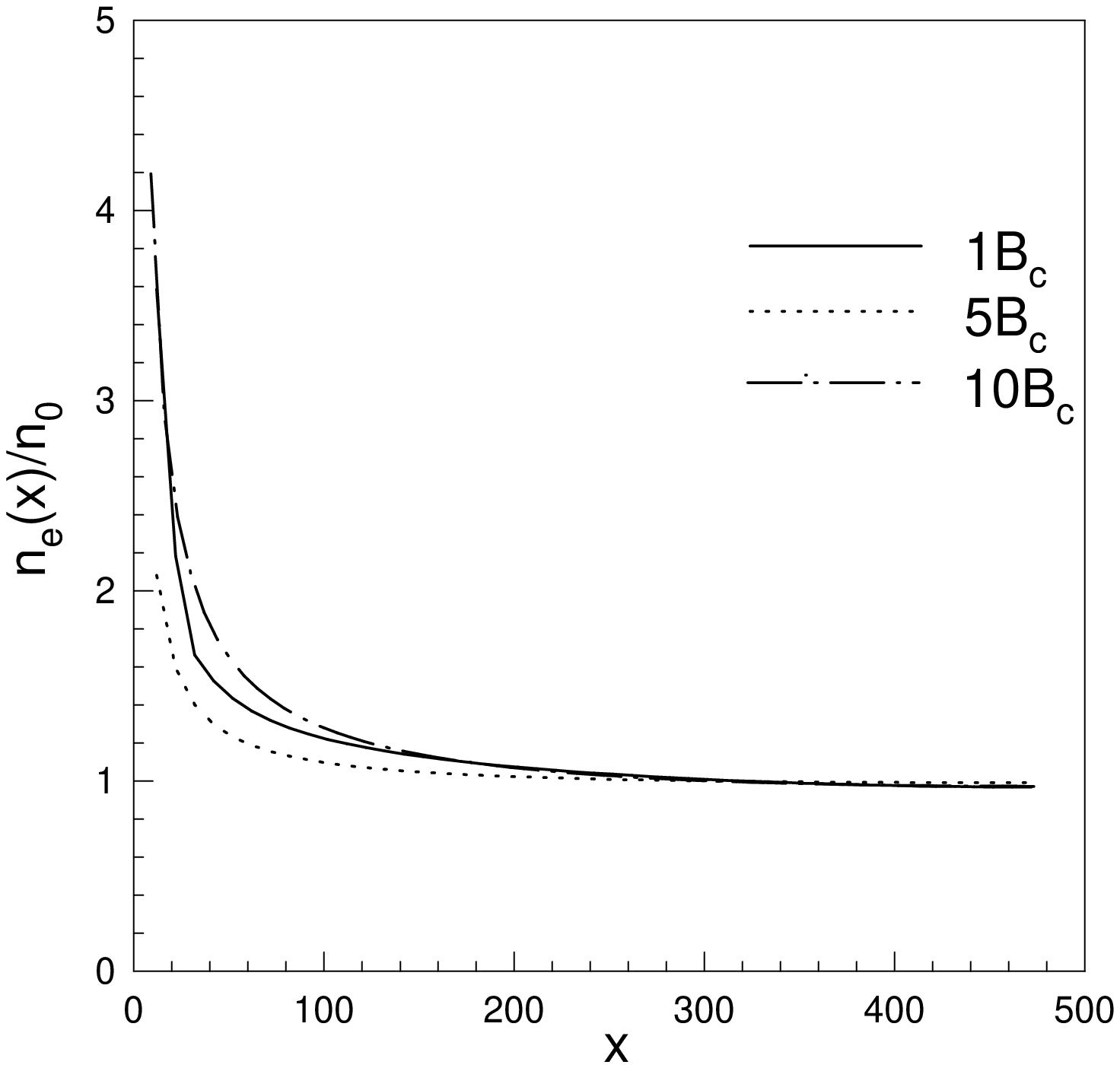,height=7.7cm,width=7.7cm}}
\caption
{Plots for $n_e({\rm x})/n_0$ as functions of dimensionless radial co-ordinate ${\rm x}$ at density $1.53\times 10^7$ gm cm$^{-3}$ for $^{12}$C.}
\label{fig3}
\vspace{4.0cm}
\end{figure}
\noindent 

\begin{figure}[hb!]
\vspace{0.0cm}
\eject\centerline{\epsfig{file=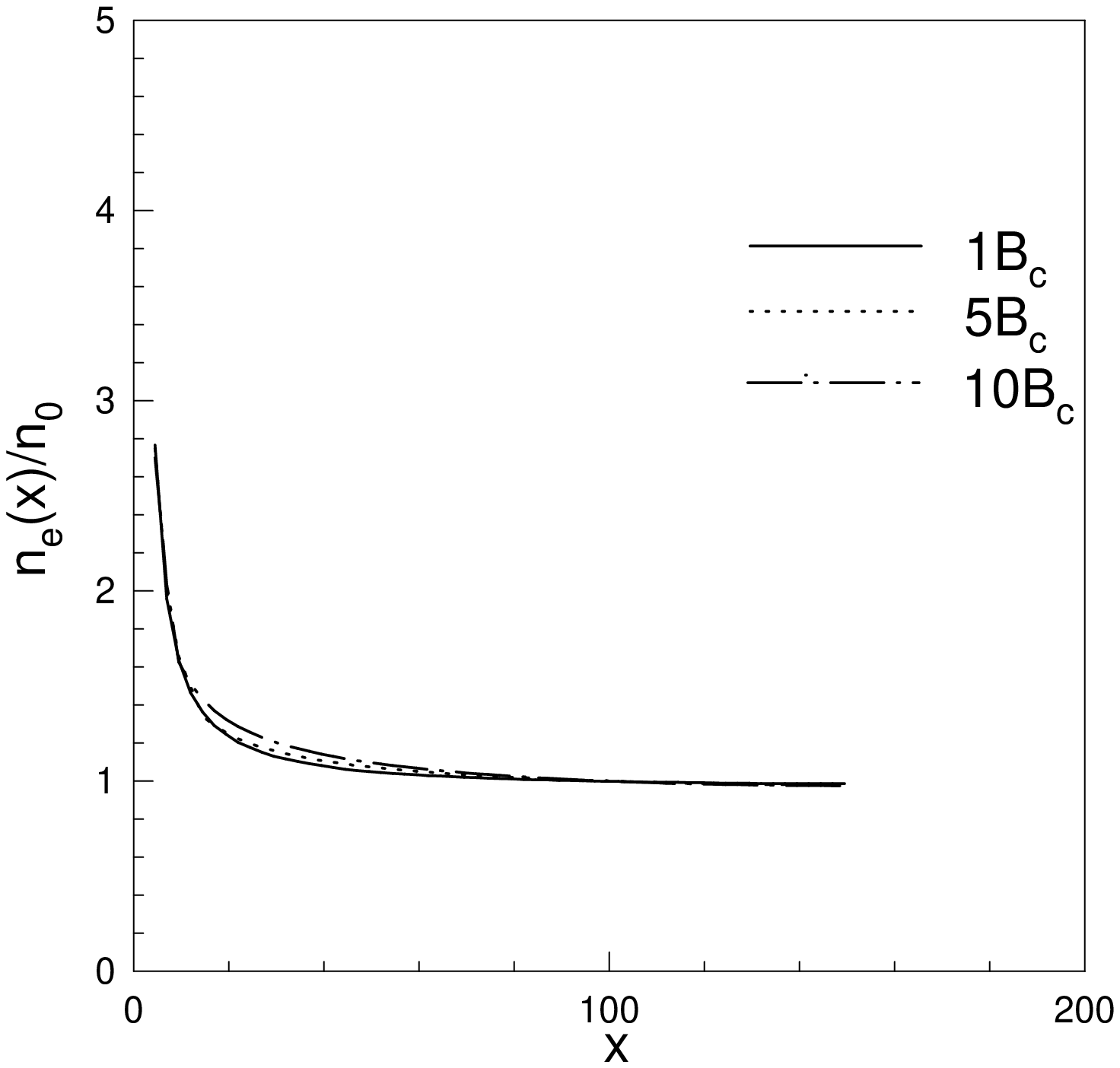,height=7.7cm,width=7.7cm}}
\caption
{Plots for $n_e({\rm x})/n_0$ as functions of dimensionless radial co-ordinate ${\rm x}$ at density $4.96\times 10^8$ gm cm$^{-3}$ for $^{12}$C.}
\label{fig4}
\vspace{0.0cm}
\end{figure}
\noindent 

\begin{figure}[ht!]
\vspace{0.0cm}
\eject\centerline{\epsfig{file=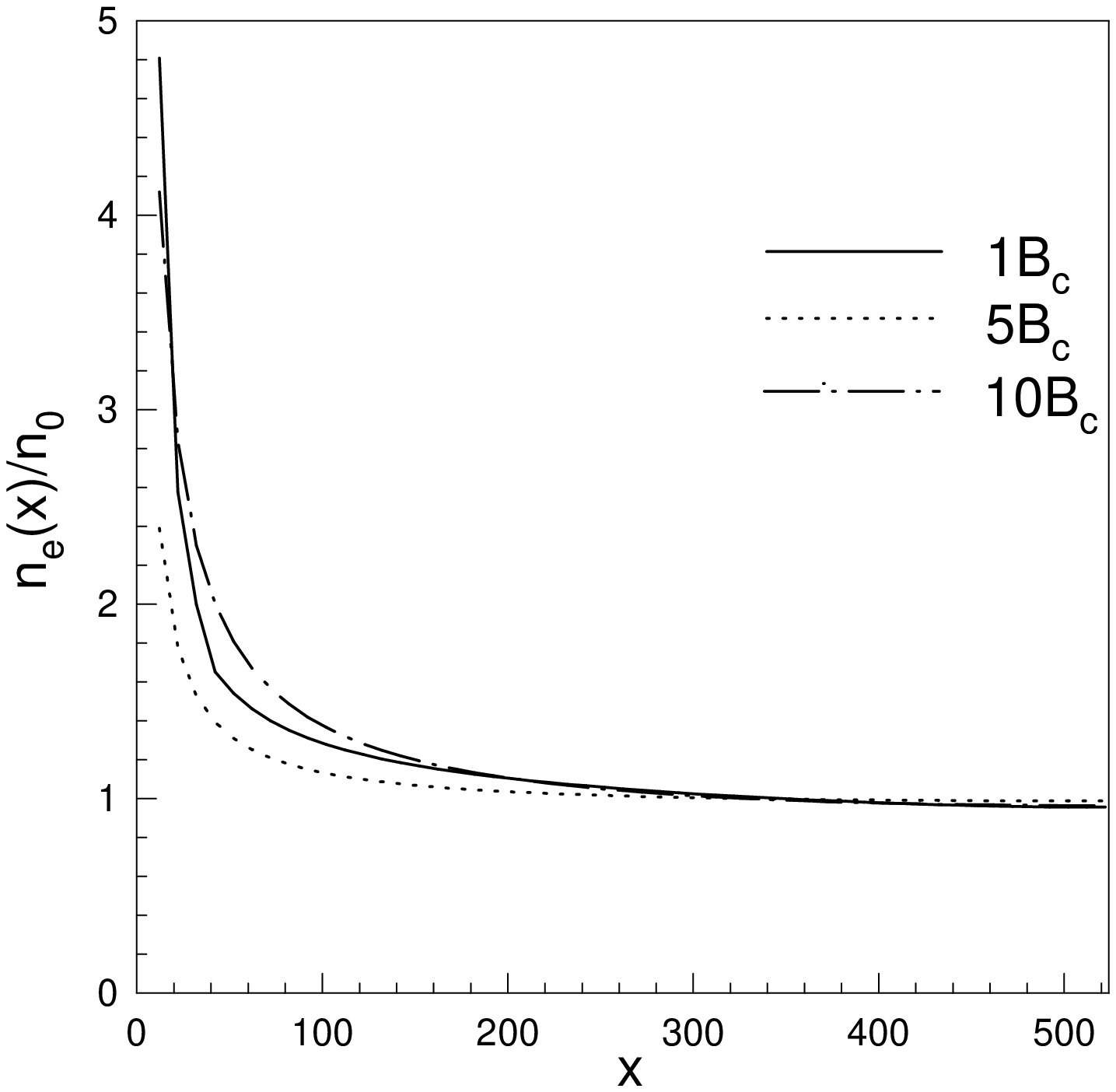,height=7.7cm,width=7.7cm}}
\caption
{Plots for $n_e({\rm x})/n_0$ as functions of dimensionless radial co-ordinate ${\rm x}$ at density $1.53\times 10^7$ gm cm$^{-3}$ for $^{16}$O.}
\label{fig5}
\vspace{4.0cm}
\end{figure}
\noindent 

\begin{figure}[hb!]
\vspace{0.0cm}
\eject\centerline{\epsfig{file=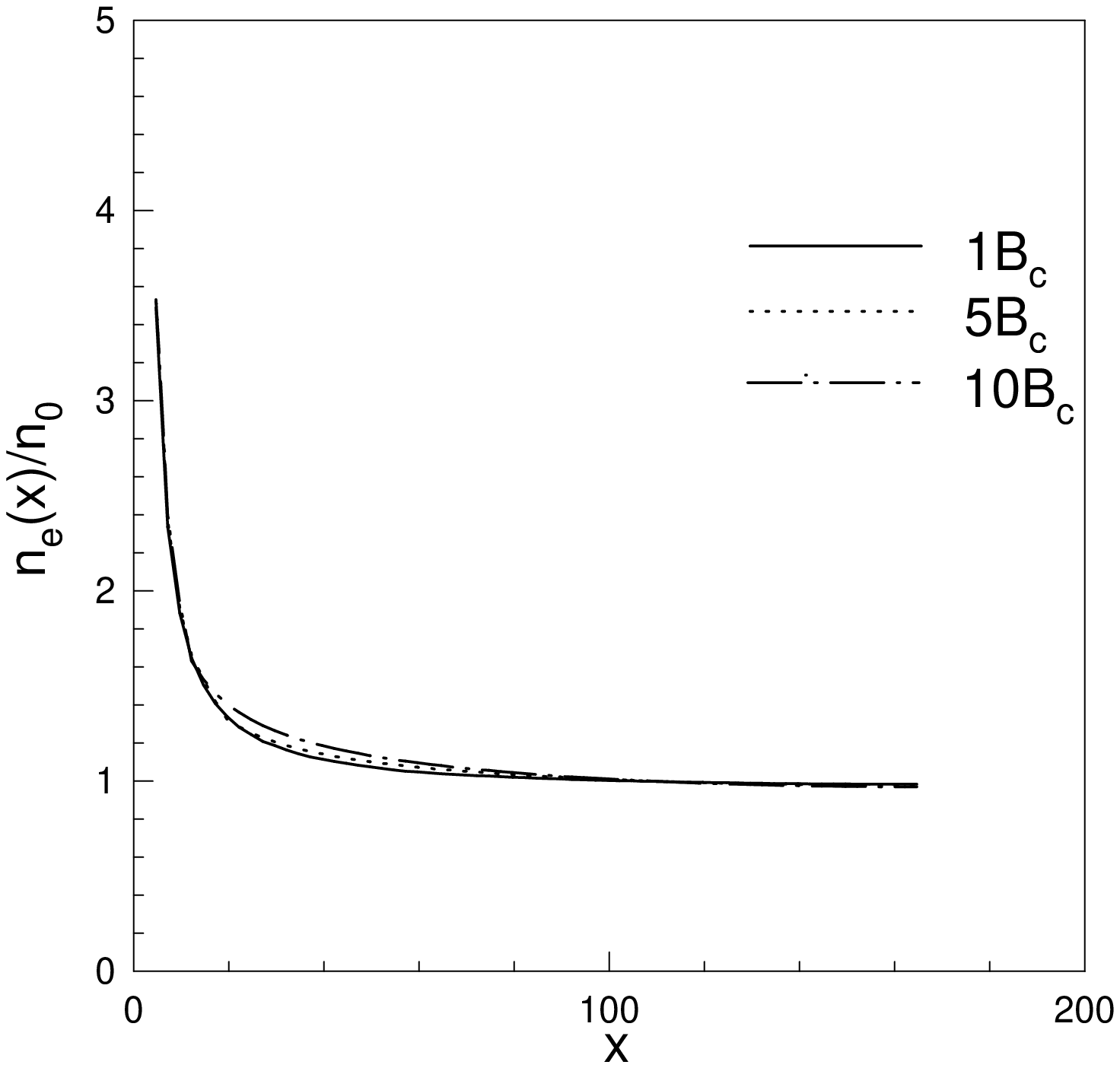,height=7.7cm,width=7.7cm}}
\caption
{Plots for $n_e({\rm x})/n_0$ as functions of dimensionless radial co-ordinate ${\rm x}$ at density $4.96\times 10^8$ gm cm$^{-3}$ for $^{16}$O.}
\label{fig6}
\vspace{0.0cm}
\end{figure}
\noindent 

\begin{figure}[ht!]
\vspace{0.0cm}
\eject\centerline{\epsfig{file=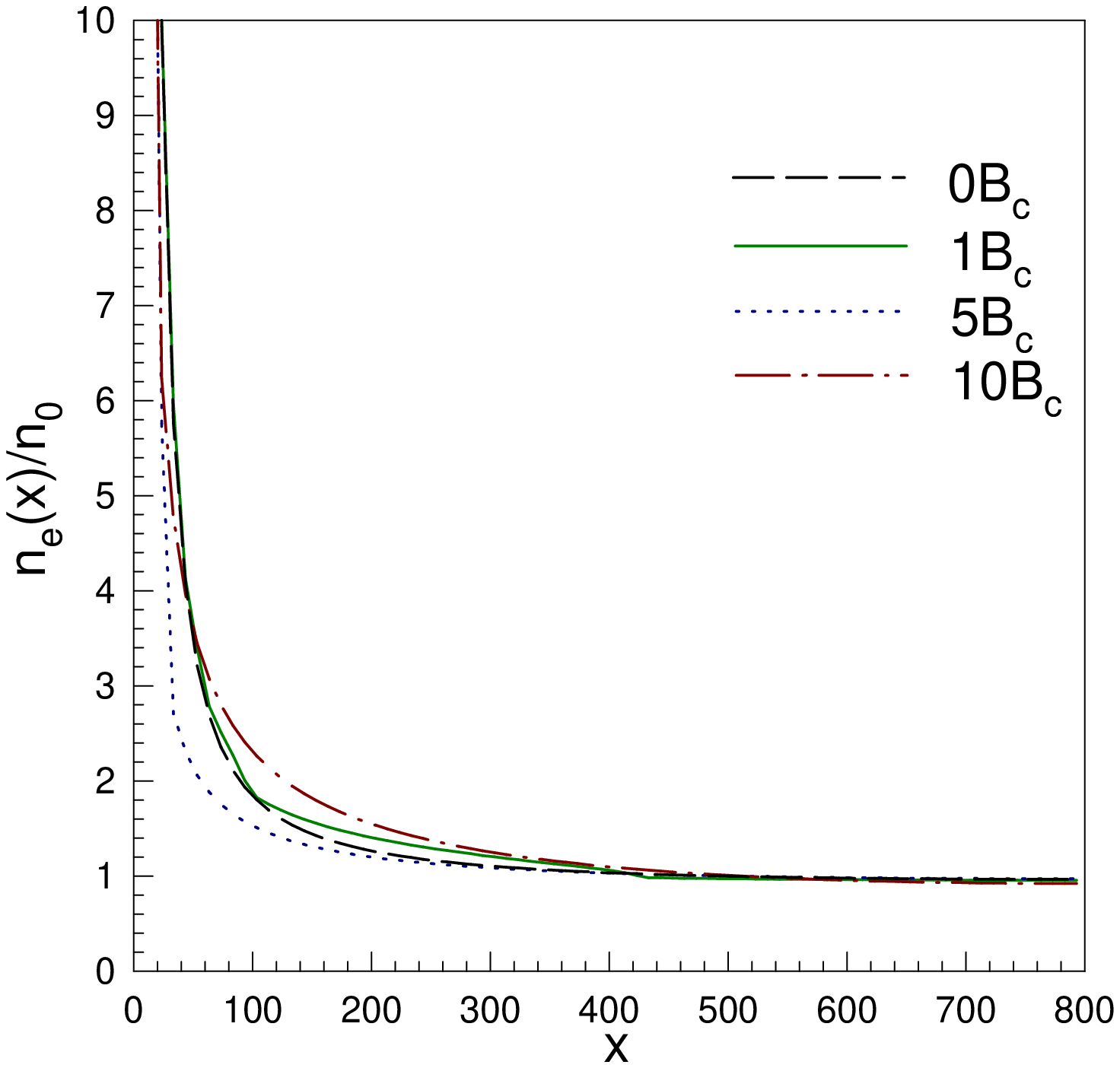,height=7.7cm,width=7.7cm}}
\caption
{Plots for $n_e({\rm x})/n_0$ as functions of dimensionless radial co-ordinate ${\rm x}$ at density $1.53\times 10^7$ gm cm$^{-3}$ for $^{56}$Fe.}
\label{fig7}
\vspace{4.0cm}
\end{figure}
\noindent 

\begin{figure}[hb!]
\vspace{0.0cm}
\eject\centerline{\epsfig{file=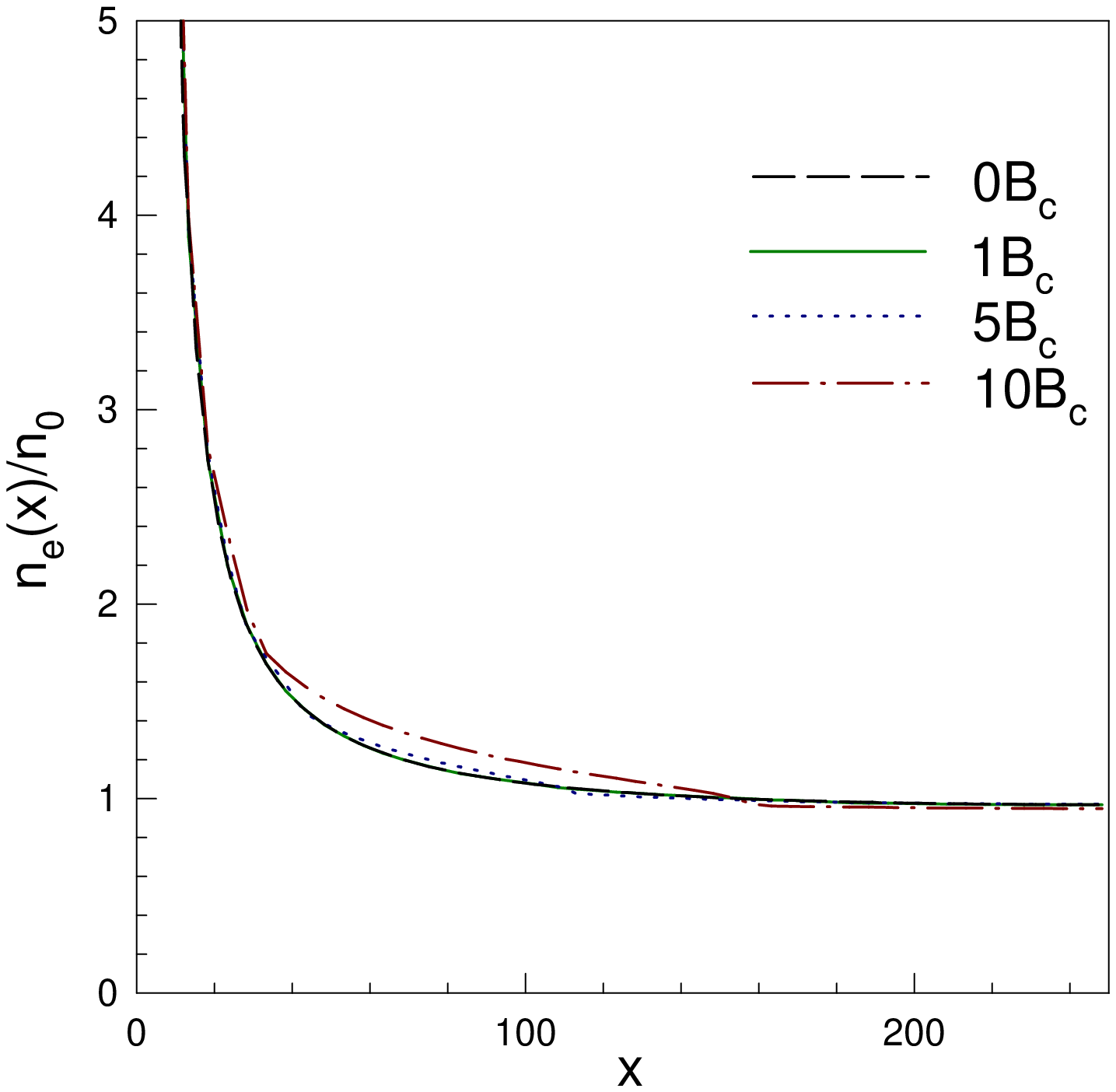,height=7.7cm,width=7.7cm}}
\caption
{Plots for $n_e({\rm x})/n_0$ as functions of dimensionless radial co-ordinate ${\rm x}$ at density $4.96\times 10^8$ gm cm$^{-3}$ for $^{56}$Fe.}
\label{fig8}
\vspace{0.0cm}
\end{figure}
\noindent 

\begin{figure}[ht!]
\vspace{0.0cm}
\eject\centerline{\epsfig{file=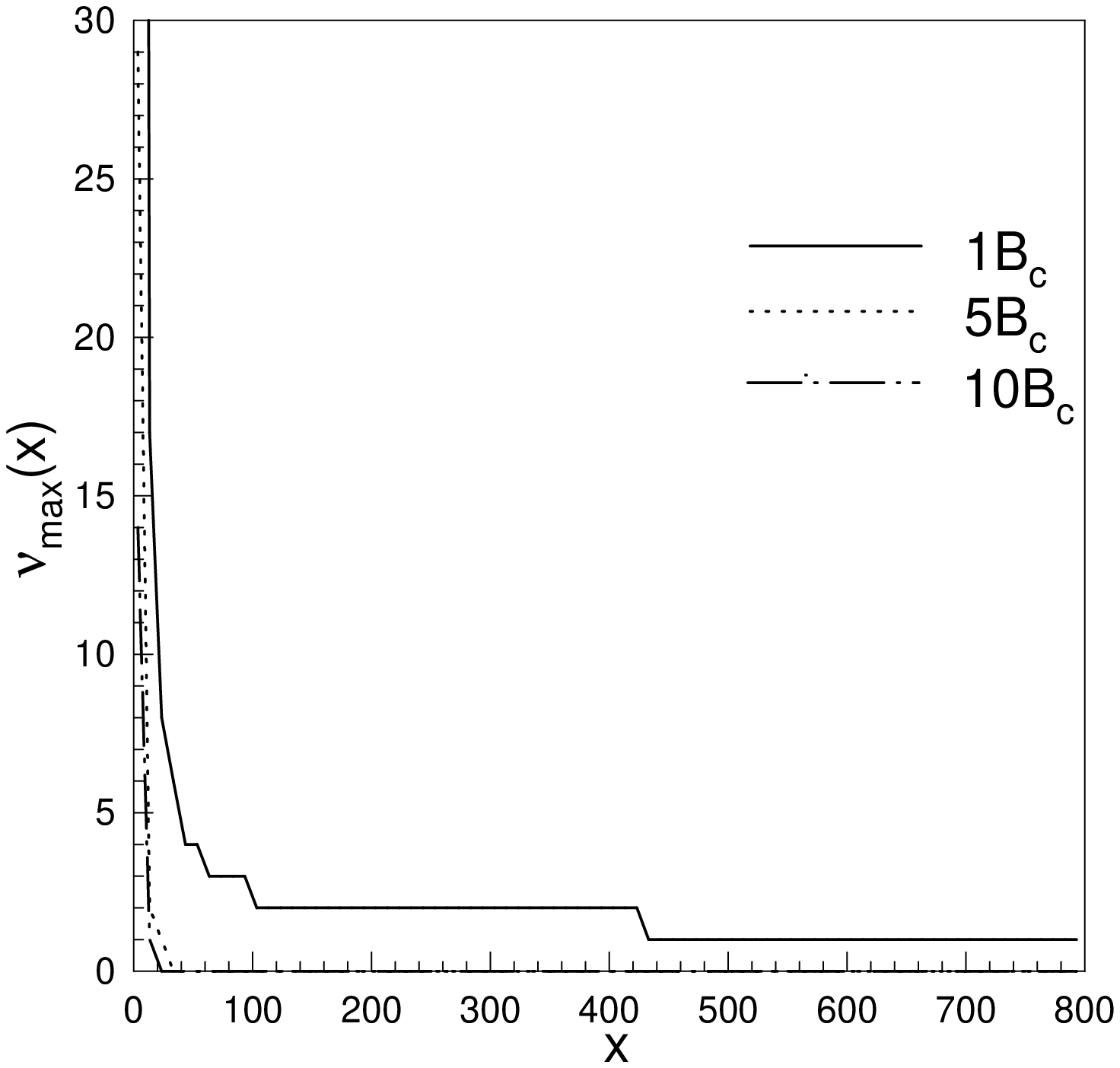,height=7.7cm,width=7.7cm}}
\caption
{Plots for $\nu_{\rm max}$ as functions of dimensionless radial co-ordinate ${\rm x}$ at density $1.53\times 10^7$ gm cm$^{-3}$ for $^{56}$Fe.}
\label{fig9}
\vspace{4.0cm}
\end{figure}
\noindent 

\begin{figure}[hb!]
\vspace{0.0cm}
\eject\centerline{\epsfig{file=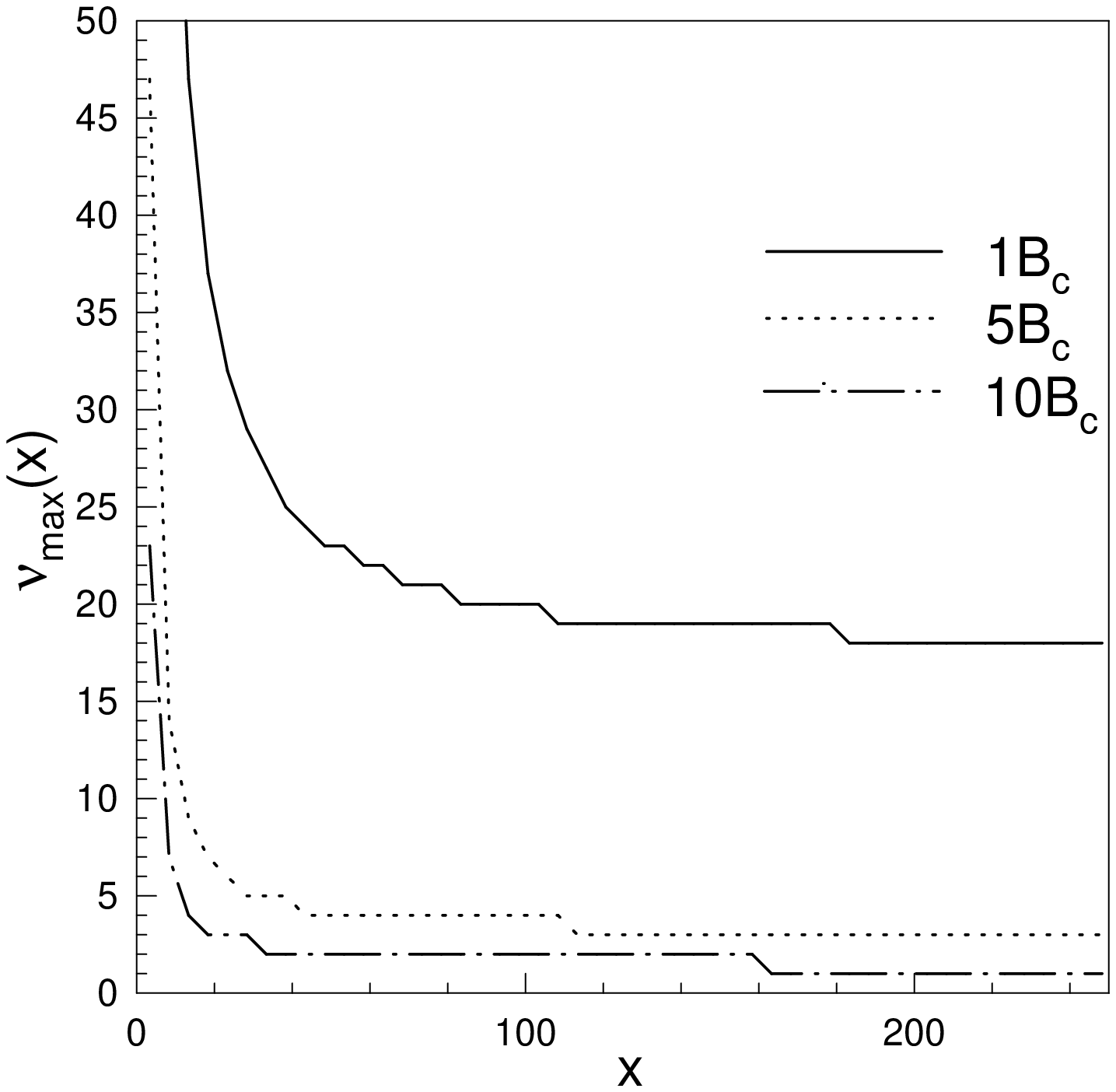,height=7.7cm,width=7.7cm}}
\caption
{Plots for $\nu_{\rm max}$ as functions of dimensionless radial co-ordinate ${\rm x}$ at density $4.96\times 10^8$ gm cm$^{-3}$ for $^{56}$Fe}
\label{fig10}
\vspace{0.0cm}
\end{figure}
\noindent              

\begin{figure}[ht!]
\vspace{0.0cm}
\eject\centerline{\epsfig{file=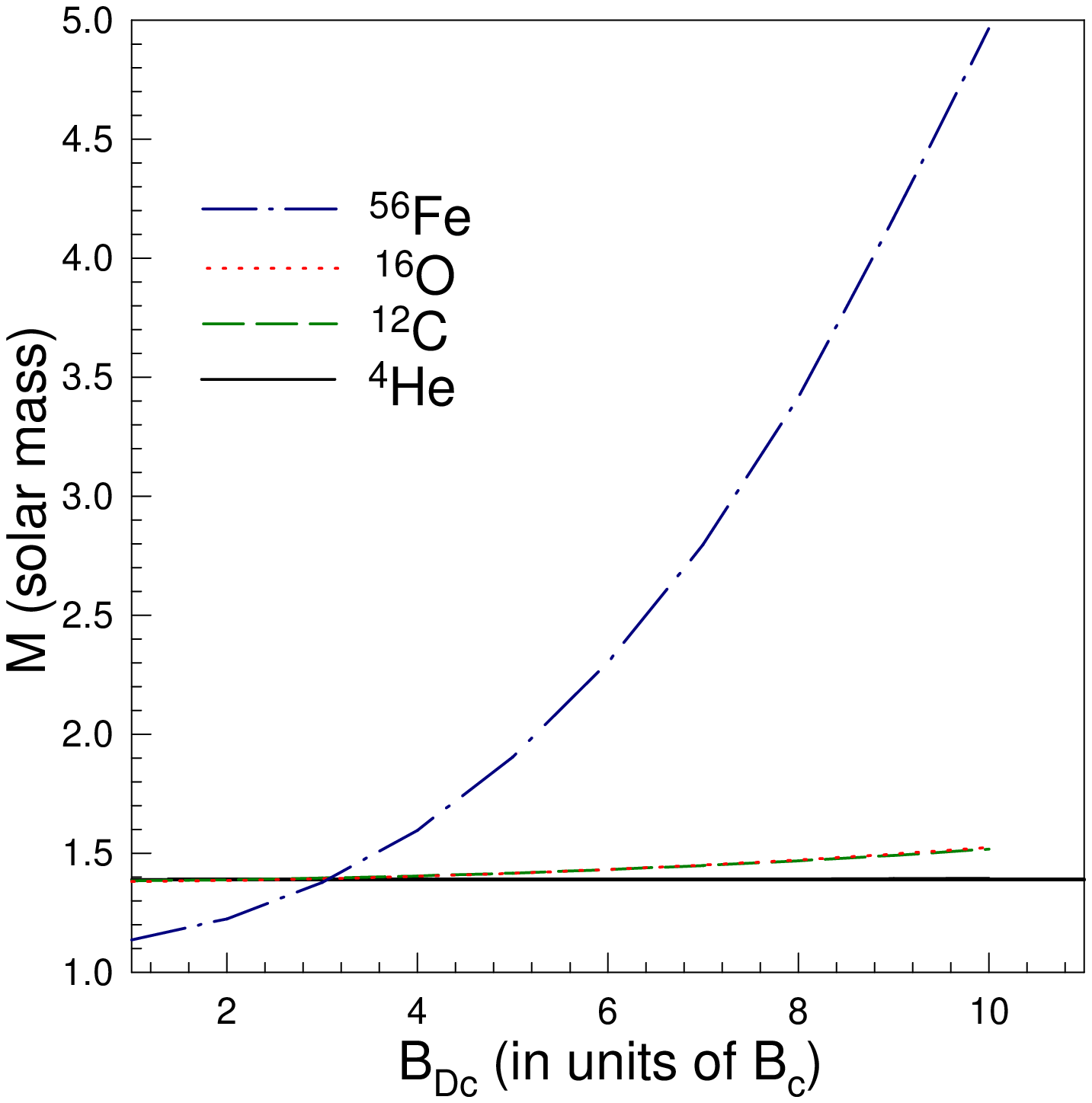,height=7.7cm,width=7.7cm}}
\caption
{Plots for maximum (critical) masses of magnetized white dwarfs as a function of central magnetic field.}
\label{fig11}
\vspace{4.0cm}
\end{figure}
\noindent 
 
\begin{figure}[hb!]
\vspace{0.0cm}
\eject\centerline{\epsfig{file=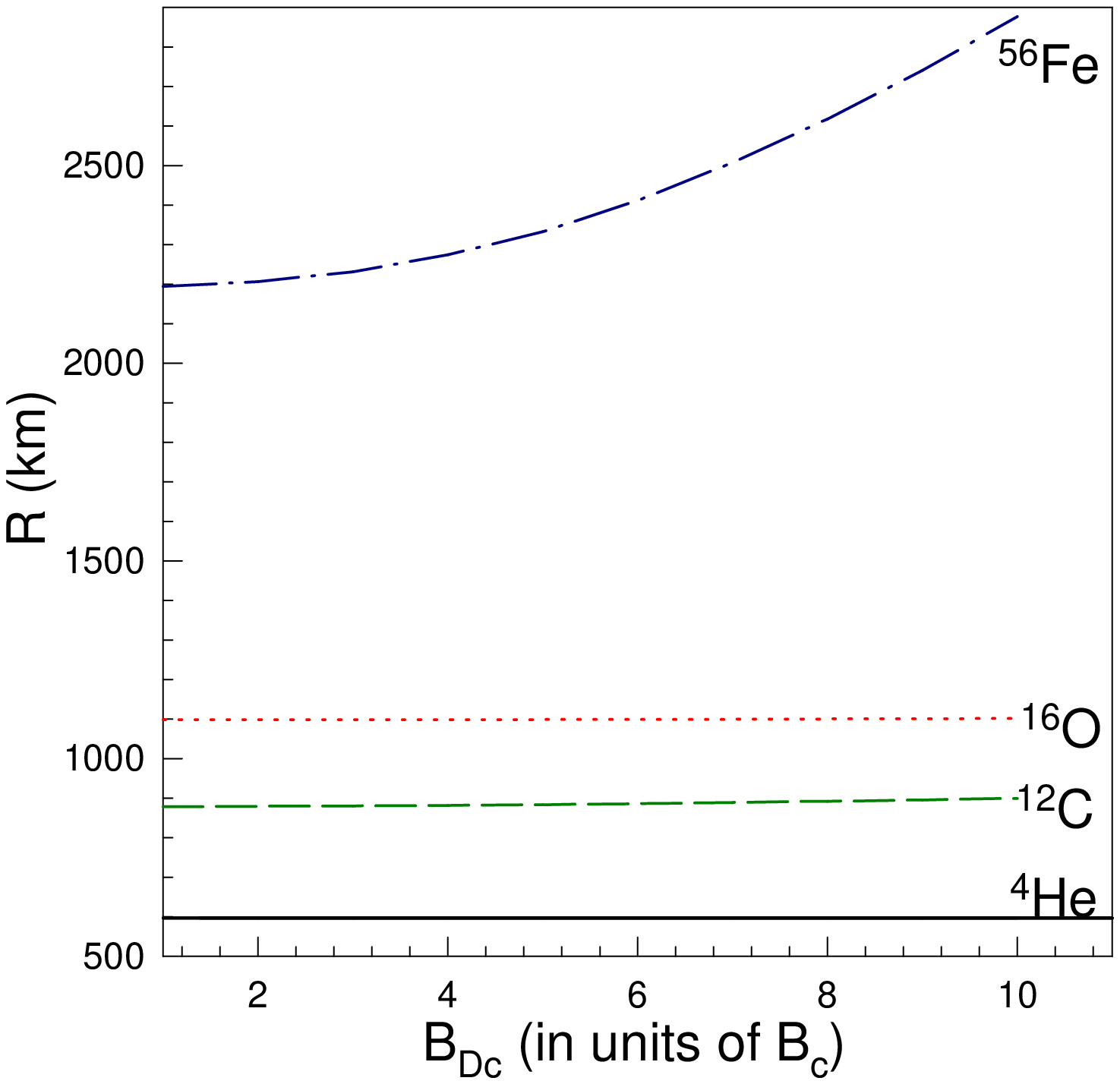,height=7.7cm,width=7.7cm}}
\caption
{Plots for radii of maximum (critical) masses of magnetized white dwarfs as a function of central magnetic field.}
\label{fig12}
\vspace{0.0cm}
\end{figure}
\noindent 

    Compact stars are believed to be formed from the stellar collapse of main sequence heavy stars. At the end of their stellar evolution when the thermal pressure becomes incapable of balancing the gravitational pull it collapses down till the Fermi degeneracy pressure of electrons grows so much that it halts the collapse. This provides the star a stable configuration, when the outward degeneracy pressure balances the inward gravitational pull, which can be yielded from the equation of hydrostatic equilibrium i.e. TOV equation \cite{TOV39a,TOV39b}. The pressure gradient $\frac{dP}{dr}$ from TOV can be intuitive to understand the configuration of such a collapsed star. As relativistically or non-relativistically the electron Fermi Degeneracy pressure contributed from the constituting matter of the star can be expressed as $P(r)=constant \times \rho(r)^\gamma$. Hence the $\rho=constant$ cannot be a possible solution for the interior of such a star made up of highly compressed degenerate matter. On the other hand as every term in the right hand side of the pressure gradient equation is positive, $\frac{dP}{dr}$ must be $\leq 0$. The equality holds for the matter at infinity i.e. on the surface of the star as is obvious from the equation. Hence the TOV equation governs a configuration with pressure, maximum at the centre of the star, which gradually approaches 0 on the surface implying a similar density distribution. But when the star collapses it is supposed that the magnetic field of the progenitor is trapped into it. So the magnetic field of the collapsed star is nothing but how the flux is getting conserved in the new configuration. Hence to preserve the flux conservation the magnetic field of the new configuration can be connected with the progenitor's one by the relation $\frac{B}{B_{prog}}= \left(\frac{\rho}{\rho_{prog}}\right)^\frac{2}{3}$ where $B (B_{prog})$ and $\rho (\rho_{prog})$ are the white dwarf's magnetic field (progenitor star) and density (progenitor star), respectively. If for simplicity we assume a constant distribution of magnetic field over the progenitor star, the surface magnetic field gets fixed in the range of $10^6-10^9$ gauss confirmed from observations \cite{Ke13,Ke15,Fe15}. As the density at the centre of white dwarf can be as high as $\sim 10^9-10^{10}$ times the density near surface, central magnetic field can go as high as $\sim 10^{12}-10^{15}$ gauss. 

\begin{table}[htbp]
\centering
\caption{The onset of inverse beta decay instability for $^4$He, $^{12}$C, $^{16}$O and $^{56}$Fe. The experimental inverse $\beta$-decay energies $\epsilon_\beta(Z)$ have been taken from Table-1 of \cite{Au03}. The corresponding critical density for the uniform electron density model $\rho^{\beta,unif}_{crit}$ is calculated using Eq.(\ref{seqn31}) while the critical density $\rho^{\beta,relFMT}_{crit}$ under magnetic field for the relativistic FMT case is calculated by the condition that $E_F$ given by Eq.(\ref{seqn22}) reaches asymptotically the inverse $\beta$ decay threshold energy $\epsilon_\beta(Z)$ whose numerical values are taken from \cite{Au77}, see also \cite{Sh83}.}
\begin{tabular}{||c|c|c|c||}
\hline 
\hline
Decay channel&$\epsilon_\beta(Z)$&$\rho^{\beta,relFMT}_{crit}$(5B$_c$)&$\rho^{\beta,unif}_{crit}$ \\ \hline
&MeV& g cm$^{-3}$ &g cm$^{-3}$ \\ \hline
\hline
$^4$He$\rightarrow$ $^3$H+n$\rightarrow$ 4n &20.596&1.368$\times$10$^{11}$&1.37$\times$10$^{11}$ \\
$^{12}$C$\rightarrow$ $^{12}$B$\rightarrow$ $^{12}$Be&13.370&3.812$\times$10$^{10}$&3.88$\times$10$^{10}$ \\
$^{16}$O$\rightarrow$ $^{16}$N$\rightarrow$ $^{16}$C &10.419&1.785$\times$10$^{10}$&1.89$\times$10$^{10}$ \\
$^{56}$Fe$\rightarrow$ $^{56}$Mn$\rightarrow$ $^{56}$Cr&3.695&1.139$\times$10$^{9}$&1.14$\times$10$^{9}$ \\ \hline
%\hline
\hline
\end{tabular}
\label{table2} 
\end{table}
\noindent     
    
    But if we take a magnetic field profile depending on the density through the equation $\frac{B_D}{B_s}=\left(\frac{\rho}{\rho_s}\right)^\frac{2}{3}$ where $B_D$ (in units of $B_c$) is the magnetic field at density $\rho$, $B_s$ (in units of $B_c$) is the surface magnetic field, it would provide a steep magnetic field. But in presence of magnetic field, as general relativistically every energy source is a source of gravitation the gravitational pull must be balanced by the combined effect of magnetic field and electron degeneracy pressure. Then at the centre of the star such magnetic field ($\sim 10^{15}-10^{16}$ gauss) would provide a huge magnetic pressure which may lead to local instabilities. If the magnetic pressure at centre turns out to be really high to dominate the gravitation field then the magnetic pressure may drive the collapsed matter to exhibit unstable convection flow therefore one can expect expulsion of such huge central magnetic field in a stable configuration. Secondly as the huge magnetic field can lead to production of particles in the system and the corresponding energy extraction can reduce the central magnetic field. Ultimately we infer that such a magnetic field profile cannot be sustained rather a magnetic field profile having flatness near the centre and surface is preferred over such a configuration.
        
    On the basis of the above discussion the actual calculations have been performed with varying magnetic field including the effects of energy density and pressure arising due to magnetic field in a general relativistic framework. The variation of magnetic field \cite{Ba97} inside white dwarf is taken to be of the form   
 
\begin{equation}
 B_D = B_s + B_0[1-\exp\{-\beta(\rho/\rho_0)^\gamma\}]
\label{seqn32}
\end{equation} 
\noindent
where $\rho_0$ is taken as $\rho$(r=0)/10 and $\beta$, $\gamma$ are constants. Once central magnetic field is fixed, $B_0$ can be determined from above equation. We choose constants $\beta=0.8$ and $\gamma=0.9$ to ensure the flatness of the field near surface and centre. Nevertheless, the magnetic field profile could have been taken with different parameter values as far as no local instability incurs. It would not alter the gross result by much unless the central magnetic field varies significantly and can be applied safely subjected to the virial condition of stability that the magnetic energy never exceeds the gravitational binding energy \cite{Ch13}. The maximum limit of central magnetic field strength of the configuration given by Eq.(\ref{seqn32}), therefore, must be chosen carefully owing to all these stability criterion and for the present calculations have been kept 10$B_c$ which is $4.414 \times 10^{14}$ gauss, near to the lower of the maximum limit as suggested by N. Chamel et al. \cite{Ch13,Ch15}. The surface magnetic field $B_s$ $\sim10^{9}$ gauss estimated by observations.

    In Figs.-1-8 plots for $n_e({\rm x})/n_0$ as functions of dimensionless radial co-ordinate ${\rm x}$ of Wigner-Seitz cell have been shown for $^4$He, $^{12}$C, $^{16}$O and $^{56}$Fe under the influence of three different magnetic fields and for two different densities where $n_0$ is number density assuming uniform distribution inside. In Figs.-9,10 plots for maximum Landau levels $\nu_{\rm max}$ as functions of dimensionless radial co-ordinate $x$ of Wigner-Seitz cell have been shown for $^{56}$Fe under the influence of three different magnetic fields and for two different densities. 

    The configurations having predominantly high nuclear binding energy are $^4$He, $^{12}$C, $^{16}$O, $^{56}$Fe with 7.08 MeV, 7.68 MeV, 7.73 MeV and 8.80 MeV per nucleon respectively. In the binding energy per nucleon versus mass number curve $^{56}$Fe is sitting at the maximum whereas the other three nuclide occupy three peaks in the curve. So in the present calculation these four types of configurations have been considered. We assumed that a progenitor star after running out of its nuclear fuel it collapses to finally form a white dwarf of either of these above mentioned materials. If the collapse can ignite further nuclear fission the white dwarf can again act as living star and go under the same process of stellar evolution till the configuration of $^{56}$Fe is reached as no further evolution is possible beyond this because this is the most stable nuclide having the maximum binding energy per nucleon.
  
    For different types of atoms of same density, the Wigner-Seitz cell radius $R_{WS} \propto A_r^\frac{1}{3}$. Hence more the atomic number more is the trapped magnetic flux for same magnetic field profile. For simplicity if one assumes a constant magnetic field distribution which after a gravitational collapse form a magnetic field profile as used in Eq.(\ref{seqn32}) then it is quite obvious that to form an Iron white dwarf, which is able to encompass a huge number of flux compared to $^4$He, $^{12}$C or $^{16}$O, a larger amount of mass of progenitor star to be collapsed than the other configurations. On the other hand it may as well be formed after many of stellar evolution cycles so that it may be able to hold such kind of a magnetic field profile consistent with large amount of magnetic flux encompassing for Iron cell. This factor in turn reflects a possibility of getting considerably higher critical masses for Iron white dwarf than that of Helium, Carbon or Oxygen white dwarfs. It can be envisaged that higher the central magnetic field higher the trapped magnetic flux implying larger collapse of the progenitor. Hence white dwarfs can be predicted with larger critical masses with increasing magnetic field. These factors can be confirmed from the figures. In Figs.-11,12 masses and radii of magnetized white dwarfs  are shown as functions of central magnetic field. Present calculations estimate that the effect is minimal for a $^4$He white dwarf and exceeds slightly the Chandrasekhar's limit for $^{12}$C and $^{16}$O white dwarfs while for $^{56}$Fe the maximum mass can be as high as $\sim$5 $M_\odot$. These results are quite useful in explaining the peculiar, overluminous type Ia supernovae that do not conform to the traditional Chandrasekhar mass-limit. 

\noindent
\section{ Summary and conclusion }
\label{section6}
    
    In summary, we have considered each compressed atom to be confined within a Wigner-Seitz cell with nucleus at its centre surrounded by relativistic degenerate electrons. The effect of strong magnetic field is incorporated by modifying the density of states of electrons due to Landau quantization. The Coulomb screening substantially alters the number density of electrons from the uniform one which reduces the effective degeneracy pressure at the cell boundary causing modification of EoS of the white dwarf matter. Any source of energy gravitates according to general relativity. Hence, the presence of magnetic field contributes additively to energy density and pressure. The masses and radii of magnetized white dwarfs, with a density dependent magnetic field profile which is high at the centre, have been calculated by solving the TOV equations. We find that the critical mass limit, subjected to the stability against inverse $\beta$ decay, of such white dwarfs increase with the magnitude of the central magnetic field. However, this increase in critical mass depends on the mass number of atom. The effect is minimal for a $^4$He white dwarf and exceeds largely the Chandrasekhar's limit for $^{56}$Fe white dwarfs which can be as high as $\sim$5 $M_\odot$.   
                    
    To date there are about $\sim$250 magnetized white dwarfs with well determined fields \cite{Fe15}
and over $\sim$600 if objects with no or uncertain field determination \cite{Ke13,Ke15} are also included. Surveys such as the SDSS, HQS and the Cape Survey have discovered these magnetized white dwarfs. The complete samples show that the field distribution of magnetized white dwarfs is in the range 10$^3$-10$^9$ gauss which basically provides the surface magnetic fields. However, the central magnetic field strength, which is presumably unobserved by the above observations, could be several orders of magnitude higher than the surface field. In fact, it is the central magnetic field which is crucial for super-Chandrasekhar magnetized white dwarfs. However, the softening of the EoS accompanying the onset of electron captures and pycnonuclear reactions in the core of these stars can lead to local instabilities which set an upper limit to the magnetic field strength at the centre of the star, ranging from 10$^{14}$-10$^{16}$ gauss depending on the core \cite{Ch13} composition.    

%\newpage             

\end{document}